\documentclass[aps,prb,groupedaddress,twocolumn]{revtex4-2}

 \usepackage{graphicx,bm,amssymb}
\usepackage[dvipsnames]{xcolor} 
\usepackage{fontawesome}

\usepackage{amsmath}
\DeclareMathOperator\arctanh{arctanh}
\newcommand{\bea}{\begin{eqnarray}}
\newcommand{\eea}{\end{eqnarray}}
\newcommand{\be}{\begin{equation}}
\newcommand{\ee}{\end{equation}}
\newcommand{\eps}{\varepsilon}

\begin{document}

\title{Phase transitions and polarization switching  in quasi-one-dimensional organic ferroelectrics of phenyltetrazole family}

\author{A.P. Moina}
\email[]{alla@icmp.lviv.ua}
\affiliation{Yukhnovskii Institute for Condensed Matter Physics, 1 Svientsitskii St., 79011, Lviv, Ukraine}

\date{\today}

\begin{abstract}
Pseudospin models are proposed for description of the phase transitions, dielectric characteristics, and polarization switching in two crystals of the phenyltetrazole family. One of them, APHTZ, is a canted ferroelectric, whereas the other, MPHTZ is a simple antiferroelectric.  In APHTZ the electric field, applied perpendicularly to the axis of spontaneous polarization,  flips the polarization in one of the two sublattices, effectively rotating the non-zero net polarization by 90$^\circ$ and switching the system between two different ferroelectric configurations. The temperature-electric field phase diagrams are constructed. The diagram topology appears to be typical for the Ising-like antiferroelectric systems. 

\end{abstract}


\maketitle


\section{Introduction}

There is currently much interest \cite{Horiuchi:2025} in search for lead-free organic ferroelectric and antiferroelectric systems as promising  technologically advantageous candidates
for various applications, like high-energy-storage capacitors.

It has been very recently discovered  \cite{Horiuchi:2023} that 
 organic  crystals of the 5-Phenyl-1\textit{H}-tetrazole (PHTZ) family exhibit  ferroelectric (FE), antiferroelectric (AFE), or phase mixture FE/AFE or AFE/AFE behavior. 
Amongst several studied crystals of this family, two,  5-(4-methylphenyl)-1\textit{H}-tetrazole (MPHTZ) and 5-(4-aminophenyl)-1\textit{H}-tetrazole (APHTZ) were found to
have a long-range order. APHTZ appears to be a canted ferroelectric, with FE ordering along the $a$ axis that coexists with the AFE ordering along the $b$ axis \cite{Horiuchi:2023}. MPHTZ is a simple uniaxial AFE. The ordered phases are stable at least up to 410~K.

\begin{figure}[hbt]
	\centerline{
		\includegraphics[width=0.8\columnwidth]{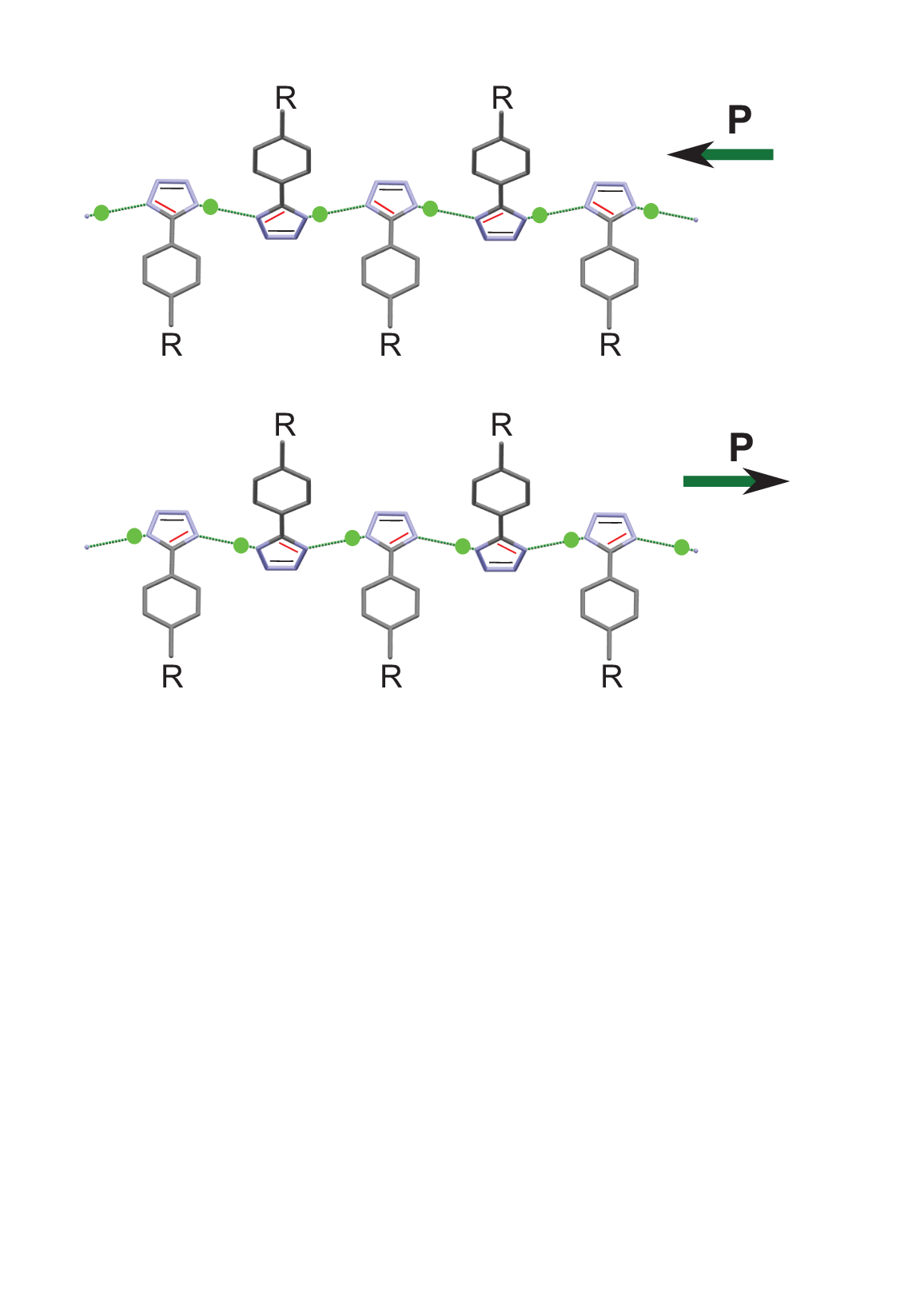}	
	}
	\caption{Two fully ordered hydrogen bonded 5-(4-\textit{R}-Phenyl)-1\textit{H}-tetrazole chains with opposite polarizations, indicated by large green arrows. Switchable and non-switchable $\pi$-bonds of tetrazole groups are shown in red and black, respectively. } \label{PHTZ-chain}
\end{figure}

The common feature and the building blocks of the structure of all PHTZ crystals are pentagonal heterocyclic tetrazole (CN$_4$) groups, hydrogen bonded into quasi-one-dimensional chains, as shown in fig.~\ref{PHTZ-chain}. The radicals $R$, linked to the tetrazole groups via phenyl rings, are different in different crystals:
$R={}$NH$_2$ in APHTZ and CH$_3$ in MPHTZ. The crystals with $R={}$Cl and $R={}$F exhibit the mixture of FE/AFE ordering due to the nearly degenerate ground states with very close energies of the FE and AFE configurations \cite{Horiuchi:2023}. The parent PHTZ compound ($R={}$H) has been identified as an AFE/AFE phase mixture, where different AFE configurations have close energies.

Formation of the dipole moments in PHTZ crystals and their switching by external electric field are a perfect example of 
proton tautomerism. Protons on the hydrogen bonds move in double-well potentials. The position of a switchable double electron bond (the $\pi$-bond between one of the two N--C pairs of the tetrazole group) is determined by the position of the proton on the linking hydrogen bond. As illustrated in fig.~\ref{PHTZ-chain}, the relocation of the proton to the other site along the hydrogen bond switches the $\pi$-bond to the other N--C pair in the group.  The sublattice polarization, according to the Berry phase calculations \cite{Horiuchi:2023}, is almost equally formed by the switchable $\pi$-bonds and by the ionic displacements.

Switching of a sublattice polarization by the external electric field and the resulting transition from the AFE to FE phase is a signature feature of antiferroelectrics. In uniaxial AFEs the polarization flips by 180$^\circ$ at once, which is revealed as a classical double $P$-$E$ hysteresis loop. In multiaxial AFEs, like  squaric acid \cite{Horiuchi:2018,Moina:2021,Moina:2021:2} or CPPLA \cite{Horiuchi:2021} the loops are quadruple, because of the two-step polarization rotation. MPHTZ, being a uniaxial AFE,  exhibits a classical switching process \cite{Horiuchi:2023}. APHTZ, being a canted FE, possesses compensated dipole moments along the $b$ axis. Thus, it potentially should also exhibit polarization switching: the electric field applied along the $b$-axis can flip the dipoles in one of the sublattices, inducing a transition to the state, where the ordering is FE along the $b$ axis and AFE along the $a$ axis. In APHTZ one thus expects to observe a single FE loop for the field along the $a$ axis and a double loop for the field applied along the $b$ axis.

Theoretical model description of the phase transitions and polarization switching in the systems with proton tautomerism is often based \cite{Moina:2021,Moina:2021:2,Moina:2020} on a certain version of the pseudospin proton ordering model. Since the pioneering works by Blinc \cite{Blinc:1960} and de Gennes \cite{deGennes:1963}, similar Ising-type models, possibly modified by inclusion or neglecting 
proton tunneling, coupling to phonons or to macroscopic lattice strains, have been extensively and successfully used to  describe the temperature, external mechanical stress or electric field behavior of the static dielectric, piezoelectric, elastic, thermal, as well as dynamic properties of numerous hydrogen bonded ferroics (see, e.g.~\cite{Moina:2021,Moina:2021:2,Moina:2020,Vaks:1975,Tokunaga:1970,Zachek:1980,Stasyuk:1999,Carvalho:1978,Zachek:2014,Zachek:2017,Moina:2020} and references therein). The apparent bistability of the proton-associated dipoles in the phenyltetrazole family crystals will enable us to use the pseudospin formalism, and the quasi-one-dimensional structure of the dipole chains will make 
such calculations particularly simple. To our best knowledge, no such description for these crystals has been carried out yet. 

In the present paper we shall consider the two crystals of this family, which exhibit the long-range ordering: APHTZ and MPHTZ.
Our goal is to develop as simple as possible pseudospin models, which would reproduce the essential features of the structure, type of ordering, available experimental data for the physical characteristics of the real crystals, and which would use as few fitting parameters as possible. The processes of polarization switching in these crystals are another focus of our attention.

\section{APHTZ}

\begin{figure}[hbt]
	\centerline{
		 \includegraphics[height=0.51\columnwidth]{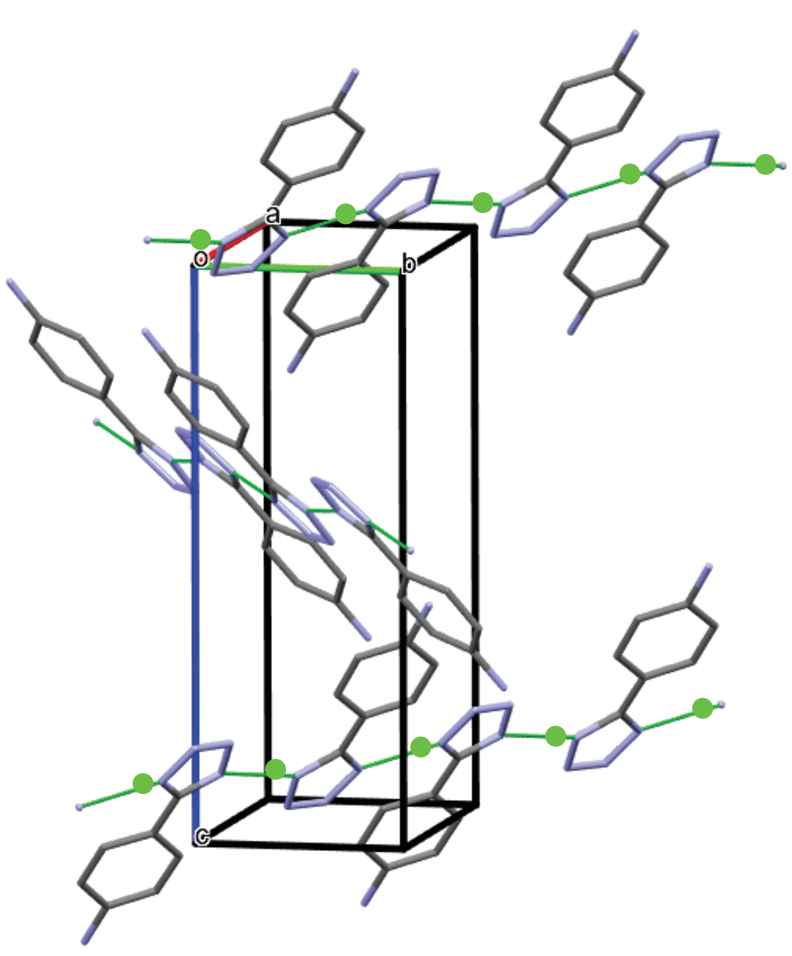}	
		 \hspace{2ex}
		\includegraphics[width=.49\columnwidth,height=.49\columnwidth]{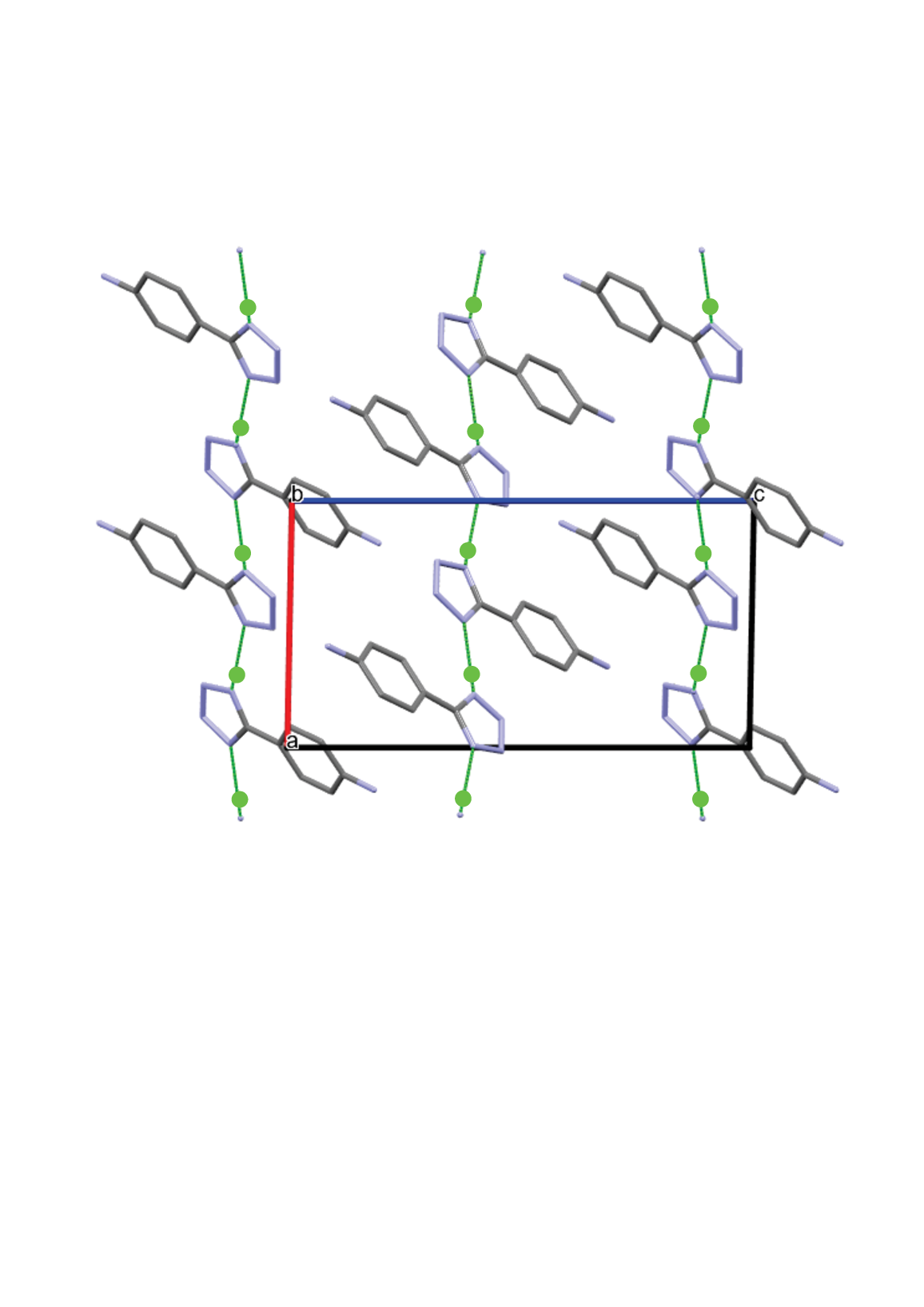}	
	}
	\caption{Two projections of the crystal structure of APHTZ, visualized \cite{Mercury} using the X-ray data of Ref.~\cite{Horiuchi:2023}. Hydrogen bonds, linking the tetrazole groups, are shown in green. The off-center displacements of protons along these bonds are exaggerated. The FE-I configuration is depicted (see text).} \label{APHTZ-struct}
\end{figure}

The monoclinic (space group $Pn$) APHTZ crystals have a layered-like structure, where each layer is formed  by the illustrated in fig.~\ref{PHTZ-chain} parallel chains of molecules, linked by hydrogen bonds. As seen in fig.~\ref{APHTZ-struct},  the layers are stacked along the $c$ axis, and the chains are alternatingly (from layer to layer) directed either along [110] or [1$\bar 1$0]. The dipole moment of each molecule thus has both $a$ and $b$ components, and different types of the long-range order along the two directions are possible.

According to the DFT calculations of Ref.~\cite{Horiuchi:2023}, the energy of the APHTZ system is lowest in the configuration, where the ordering  is FE along the $a$-axis and AFE along the $b$-axis (the shown in fig.~\ref{APHTZ-struct} fully ordered FE-I configuration \cite{Horiuchi:2023}). In this configuration, all the protons are ordered in the furthermost (along the $a$-axis) of the two equilibrium sites on each hydrogen bond, linking the tetrazole groups. The hysteresis measurements  \cite{Horiuchi:2023} also confirmed the ferroelectric ordering along the $a$-axis in APHTZ.

\begin{figure}[hbt]
	\centerline{
		\includegraphics[width=\columnwidth]{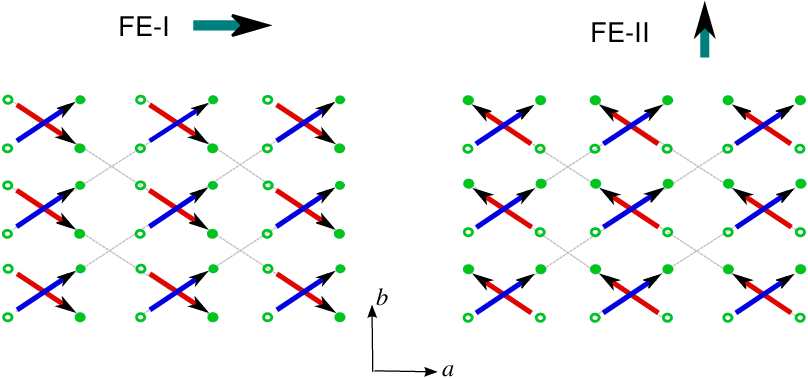}	
	}
	\caption{Fully ordered ferroelectric configurations FE-I and  FE-II of the APHTZ model (see text and Ref.~\cite{Horiuchi:2023}). Two  layers of parallel chains of dipoles oriented along [110] (blue arrows, sublattice $s=1$) and [1$\bar 1$0] (red arrows, sublattice $s=2$) are shown for each configuration. Full and open green circles are the occupied and vacant  proton  sites on each bond; each dipole always points towards the occupied site. Large green arrows indicate the net polarization directions. The  dipole sizes and relative distances between the sites, chains, etc are schematic only and do not reflect the real ones.  } \label{APHTZ-model}
\end{figure}

The proposed  model is shown in fig.~\ref{APHTZ-model}. Like the actual crystal, it consists of stacked along the $c$ axis parallel layers, each formed by parallel chains of dipoles. All chains are going along [110] in the odd layers (sublattice $s=1$) and along [1$\bar 1$0] in the even layers (sublattice $s=2$).  Each effective dipole is associated with a hydrogen bond, linking two neighboring molecules, but is formed predominantly by ionic displacements and by the switchable $\pi$-bond of the tetrazole group. Each dipole is directed towards the occupied proton site. If the proton moves from one equilibrium site on the bond to the other, the entire dipole flips by 180$^\circ$.

In the FE-I configuration (fig.~\ref{APHTZ-model}, left) all the dipoles point in the positive direction along the $a$-axis, whereas their projections on the $b$-axis alternate between the layers and are thus compensated: the system is FE along the $a$-axis and AFE along the $b$-axis.
Another possibility is the  FE-II configuration (fig.~\ref{APHTZ-model}, right), with the FE order along the $b$ axis and AFE order along the $a$ axis. The energy of this configuration was found \cite{Horiuchi:2023} to be higher than that of FE-I. It is expected that the electric field $E_b$ can induce a switching from phase FE-I to phase FE-II.

Due to the obvious bistability of each ordering element of the model (the proton in a double-well potential and an electric dipole, rigidly tied to the proton), the pseudospin formalism is most preferable for the calculations. Motion of a proton  is described by an Ising pseudospin, whose two eigenvalues $\sigma=\pm 1$ are assigned to the two equilibrium proton sites  on the bond. We assume that $\sigma=1$, if the proton sits at the right site (along the $a$ axis), and the associated dipole points to the right, and 
$\sigma=-1$ if the proton sits at the left one, and the dipole points to the left (see fig.~\ref{APHTZ-model}).

The model Hamiltonian includes pairwise long-range interactions between the dipoles, as well as the energy of their interactions with external electric fields $E_a$ and $E_b$
\begin{eqnarray}
&& 
\label{APHTZ-H}
H=-\frac12\sum_{s,s'=1,2}\sum_{ff'} J_{sf \atop s'f'}\frac{\sigma_{sf}}2\frac{\sigma_{s'f'}}2 \\
&& {}-\mu_aE_a \sum_{s=1,2}\sum_{f} \frac{\sigma_{sf}}2
-\mu_bE_b \left[\sum_{f\in 1} \frac{\sigma_{1f}}2 -
\sum_{f\in 2} \frac{\sigma_{2f}}2\right] .\nonumber
\end{eqnarray}
The summations go over two sublattices ($s=1,2$) and sites $f$ belonging to the sublattice $s$ ($f$ is a combined index, which includes the indices of the layer in the sublattice, of the chain within the layer, and of the site within the chain).

The simplest mean field approximation shall be used for  Hamiltonian (\ref{APHTZ-H}).
Considering the translational symmetry of the model,
there will be two different order parameters, one for each of the sublattices 
\[\langle \sigma_{sf} \rangle=\eta_s.\]
In absence of an external field, $\eta_1 = \eta_2$ for the FE-I configuration and $\eta_1 = -\eta_2$  for the FE-II configuration.

Applying the mean field approximation leads to 
\begin{equation}
H= NH_0-\sum_{f\in 1} h_1 \frac{\sigma_{1f}}2
- \sum_{f\in 2} h_2\frac{\sigma_{2f}}2,
\label{APHTZ-MFA}
\end{equation}
where $N$ is the total number of dipoles in each sublattice;
\begin{equation}
\label{APHTZH0}
H_0=\frac18\left[J_{11}\eta_1^2+2J_{12}\eta_1\eta_2+J_{11}\eta_2^2\right],
\end{equation}
and
\begin{eqnarray}
h_{1}=\frac{J_{11}\eta_1+J_{12}\eta_2}{2}+\mu_aE_a+\mu_bE_b, \nonumber\\
\label{APHTZh1h2}
h_{2}=\frac{J_{12}\eta_1+J_{11}\eta_2}{2}+\mu_aE_a-\mu_bE_b
\end{eqnarray}
are the mean fields acting on the pseudospins of the sublattices 1 and 2, respectively.
The intra- and intersublattice interaction constants $J_{ss'}$
\begin{eqnarray}
J_{11}=\sum_{ f'\in 1 }J_{ 1f \atop 1 f'}=
\sum_{ f'\in 2}J_{ 2f \atop 2 f'},\nonumber\\
J_{12}=\sum_{ f'\in 2}J_{1f \atop 2f'}=
\sum_{ f'\in 1}J_{2f \atop 1 f'}
\end{eqnarray}
represent the total action exerted by all pseudospins of the sublattice $s'$ on each pseudospin of sublattice $s$. Depending on the sign of $J_{12}$ at positive $J_{11}$, Hamiltonian (\ref{APHTZ-MFA}) can predict either FE-I or FE-II configuration as the ground state at $E_b=0$.

Calculation of the thermodynamic potential is straightforward, yielding
\begin{equation}
\label{APHTZ-g}
g=\frac{G}N=H_0
-\frac{1}{\beta}\ln 2\cosh \frac{\beta h_1}{2}
-\frac{1}{\beta}\ln 2\cosh \frac{\beta h_2}{2},
\end{equation}
where $\beta=(k_{\rm B}T)^{-1}$. 
Minimizing it with respect to the order parameters $\eta_1$ and $\eta_2$, we obtain the following equations for them
\begin{equation}
\label{APHTZ-ordpareq}
\eta_s=\tanh\frac{\beta h_s}2,
\end{equation}
which form is typical for the mean field approximation. Equally straightforward is to find crystal polarizations
$P=-v^{-1}\frac{\partial g}{\partial E}$
 along the $a$ and $b$ axes
\begin{eqnarray*}
	P_{a}=\frac{\mu_a}{2v}(\eta_1+\eta_2)
	, \\
	P_{b}=\frac{\mu_b}{2v}(\eta_1-\eta_2)
	 ;
\end{eqnarray*}
$v=364~\AA^3$ is the crystal volume per two formula units \cite{Horiuchi:2023}.

From here we find explicit expressions for the dielectric permittivities of the system in the FE-I configuration for zero bias fields
\begin{eqnarray}
&& \eps_i=\frac1{\eps_0}\frac{\partial P_i}{\partial E_i}=\frac{\beta\mu_i^2}{2\eps_0v\varphi_i},
\end{eqnarray}
where $\eps_0$ is the dielectric permittivity of a free space, 
$i$ is either $a$ or $b$, and
\begin{equation}
\varphi_{a,b}=\frac{1}{1-\eta_0^2}-\frac{\beta(J_{11}\pm J_{12})}{4}.
\end{equation}
Here $\eta_0\equiv\eta_1=\eta_2$ is the spontaneous order parameter. In the paraelectric phase ($\eta_0=0$),  both permittivities obey the Curie-Weiss law with the Curie-Weiss temperatures
\begin{equation}
\label{APHTZ-Tc}
T_{a,b}^{\rm{CW}}=\frac{J_{11}\pm J_{12}}{4k_{\rm B}}.
\end{equation}

Before going into the numerical calculations, the following disclaimer should be made. The fitting procedure for PHTZ crystals is complicated by a severe deficit of experimental data, like the value of the transition temperature, or the temperature curves of spontaneous polarization and static permittivity, usually available for ferroelectric crystals. The chosen values of the fitting parameters are preliminary and must be adjusted when additional data become available.

First, it is obligatory that $J_{11}>0$ to ensure the FE ordering and absence of modulation within each sublattice.
The further choice of the model parameters is based on the assumption that the ground state of the system is FE-I. In this case we must put $J_{12}>0$.

The temperature and field dependent equilibrium values of the order parameters $\eta_s$ are found by direct minimization of the thermodynamic potential (\ref{APHTZ-g})
and required to obey Eqs.~(\ref{APHTZ-ordpareq}).
When both $J_{11}$ and $J_{12}$ are positive and $E_b=0$, the system behavior is governed by
two model parameters: the sum $J_{11}+J_{12}$, irrespectively of the $J_{11}$ to $J_{12}$ ratio, and the effective dipole moment $\mu_a$.  It is known \cite{Horiuchi:2023} that the ferroelectric phase is stable at least up to 410~K.  Using the Curie-Weiss temperature $T_a^{\rm{CW}}$ as an estimate for $T_c$, from Eq.~(\ref{APHTZ-Tc}) it then follows that $(J_{11}+J_{12})/k_{\rm B}$ should exceed 1640~K. We  use $(J_{11}+J_{12})/k_{\rm B}=1700$~K, which sets the transition temperature to 425~K.

The value of effective dipole moment $\mu_a=3.3\cdot 10^{-29}$~C m
is chosen to yield the spontaneous polarization 
$P_a=7.5~\mu$C/cm$^2$ at 300~K
in agreement with with the value obtained by the Berry phase calculations and with the measured value 7.0~$\mu$C/cm$^2$ of the remanent polarization at 300~K \cite{Horiuchi:2023}. Note that normal FE-type (single) hysteresis loops $P_a$-$E_a$ were observed \cite{Horiuchi:2023}.

\begin{figure}[hbt]
	\centerline{\includegraphics[width=0.47\columnwidth,height=0.47\columnwidth]{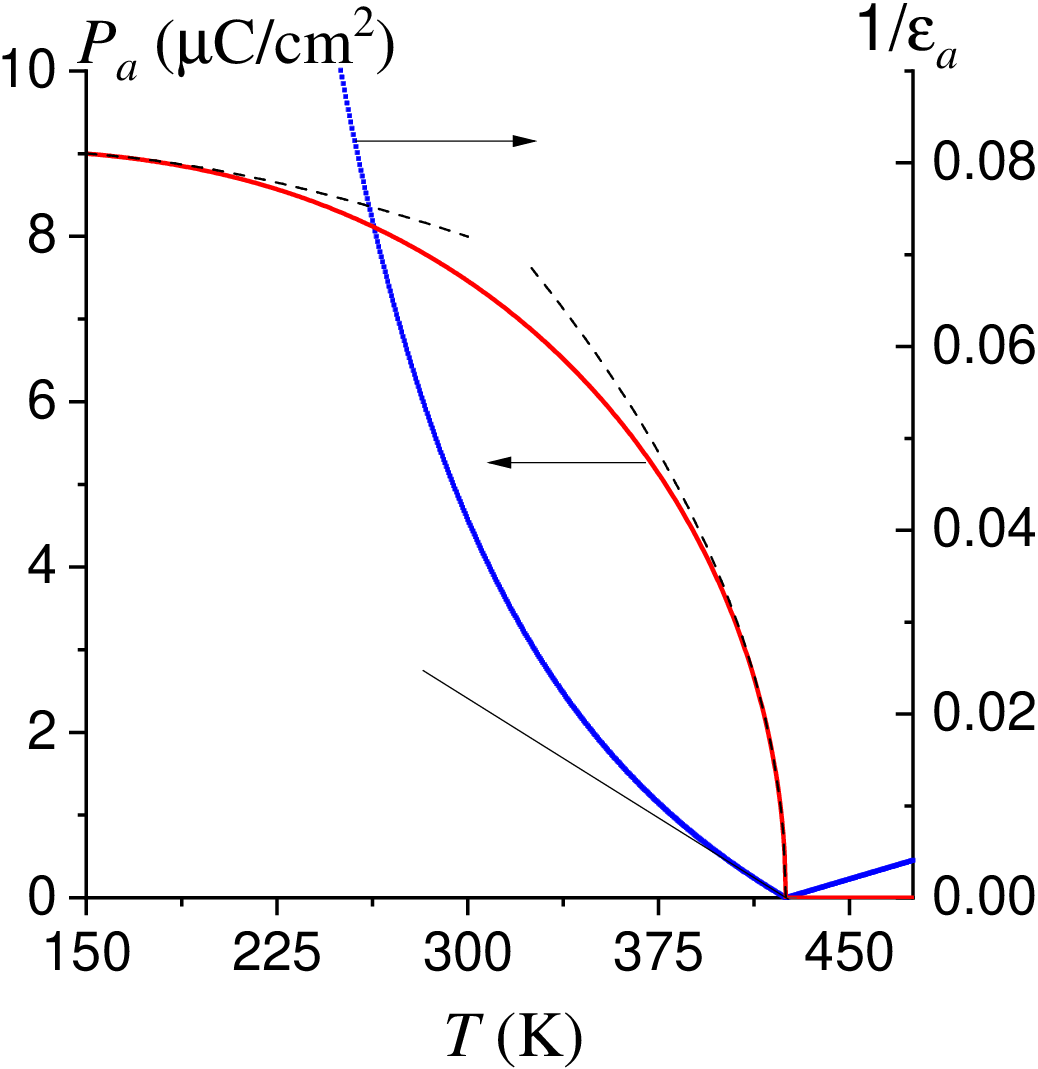}		\hspace{2ex}
		\includegraphics[width=0.47\columnwidth,height=0.47\columnwidth]{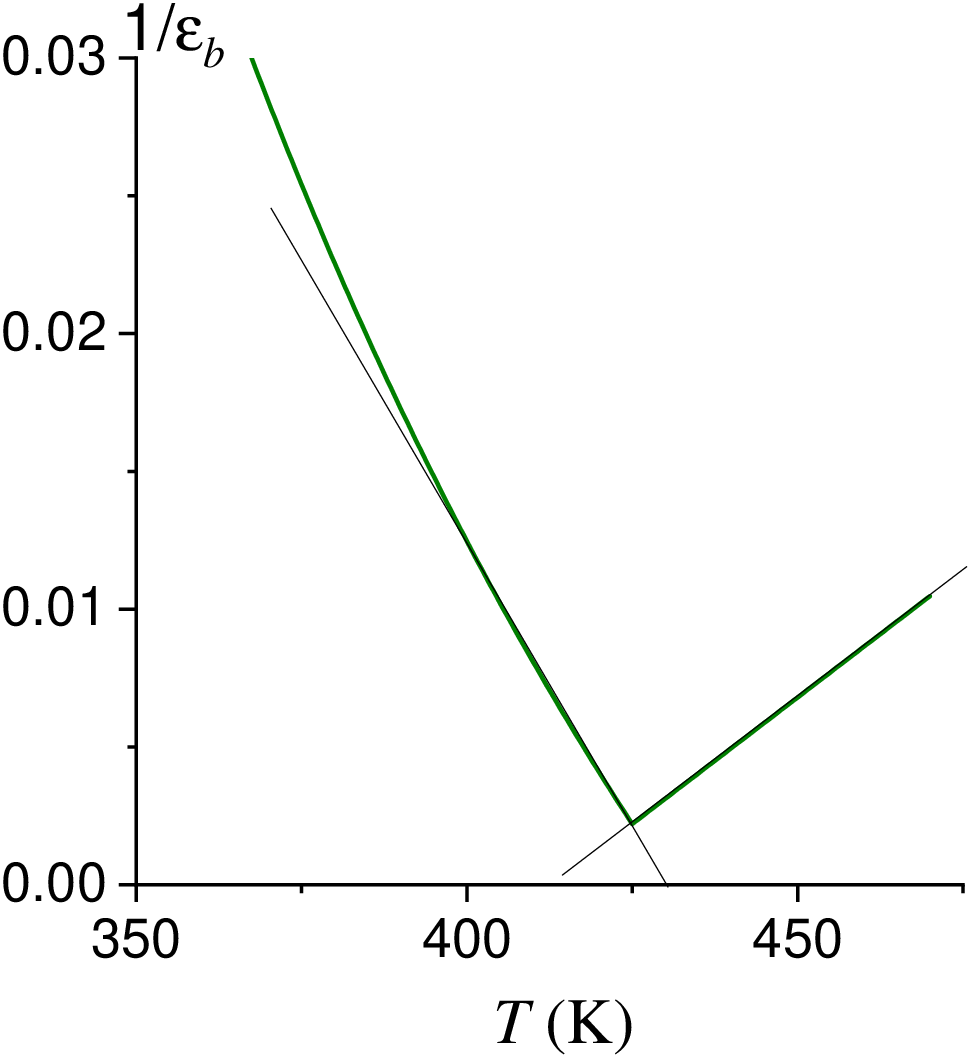}
	}
	\caption{The temperature dependences of the spontaneous polarization $P_a=
		\mu_a\eta_0/v$, inverse dielectric permittivities $\eps_a^{-1}$ (left) and  $\eps_b^{-1}$ (right) of APHTZ. Dashed lines: $P_a$ calculated with $\eta_0$ approximated according to Eqs. (\ref{sqrt}) and (\ref{etaapproxsat}).} \label{APHTZ-pol-eps}
\end{figure}

Figure~\ref{APHTZ-pol-eps} (left) shows the calculated temperature dependence of the spontaneous polarization  $P_a$ and the inverse static dielectric permittivity $\eps_a$ of the APHTZ crystal. As expected, the developed model predicts a second order phase transition, so the transition temperature $T_c$ at zero field coincides with the Curie-Weiss temperature $T_a^{\rm{CW}}$ and is given by Eq.~(\ref{APHTZ-Tc}) exactly:
\begin{equation}
T_c=\frac{J_{11}+J_{12}}{4k_{\rm B}}.
\end{equation} 

It is well known and can be easily shown by expanding the r.h.s. of Eq.~(\ref{APHTZ-ordpareq}) that at temperatures close to $T_c$
the spontaneous order parameter $\eta_0$ can be presented as
\begin{equation}
\label{sqrt}
\eta_0=\sqrt{3\frac{T_c-T}{T_c}}.\end{equation}
On the other hand, at low temperatures $\eta_0$ approaches saturation and can be  approximated as
\begin{equation}
\label{etaapproxsat}
\eta_0=1-2\exp\left(\frac{-2T_c}{T}\right).\end{equation}
As seen in fig.~\ref{APHTZ-pol-eps}, the spontaneous polarization $P_a\sim\eta_0$ follows the $(T_c-T)^{1/2}$ power law  down to about $T_c-30$~K and the exponential law, Eq.~(\ref{etaapproxsat}) at low temperatures up to about 225~K. The longitudinal permittivity follows the Curie-Weiss law both in the ferroelectric phase down to $T_c-30$~K and in the paraelectric phase. The ratio of the corresponding Curie constants is the classical 1:2. The spontaneous and associated with the field $E_a$ critical behavior of the model is thus as expected for the mean field theory of the ferroelectric Ising-like systems \cite{Lines}.

For the properties of APHTZ associated with the field $E_b$, the ratio between $J_{12}$ and $J_{11}$ becomes relevant, and so does the dipole moment $\mu_b$.
The value of $J_{12}$ can be determined from the difference between the thermodynamic potentials $g$, Eq.~(\ref{APHTZ-g}), of the fully ordered system in the FE-II and FE-I configurations ($T\to 0$, $\eta_1\to 1$, and 
$\eta_2\to -1$ or $1$, respectively), recalculated per one molecule
\[
\frac {g^{\rm{FE{-}II}}-g^{\rm{FE{-}I}}}2=\frac{J_{12}}4 =\Delta E
\]
and equated to the energy difference between the FE-II and FE-I states $\Delta E=0.53$~meV/molecule, found by the DFT calculations \cite{Horiuchi:2023}. We obtain that  $J_{12}/k_{\rm B}\approx 24$~K. Correspondingly, $J_{11}/k_{\rm B}= 1676$~K. As one can see, $|J_{12}|\ll J_{11}$, i.e. the coupling between the sublattices is much weaker than within them.

 
 The calculated  \cite{Horiuchi:2023} using the Berry phase formalism  polarization $P_a$ of the 
 FE-I configuration is about 1.5 times larger than  polarization $P_b$ of the FE-II configuration. This number is basically the ratio of the lattice constants $a/b$ \cite{Horiuchi:2023}, and it reflects the fact that the sublattice polarization vector is directed along the chains of the hydrogen bonds, i.e. along [110] or [1$\bar 1$0]. In our calculations we adopt the same ratio for $\mu_a/\mu_b$, resulting in $\mu_b=2.2\cdot 10^{-29}$~C m.

Unlike $\eps_a^{-1}$, the inverse transverse permittivity $\eps_b^{-1}$ shown in fig.~\ref{APHTZ-pol-eps} (right) 
remains finite at the transition point. The corresponding paraelectric Curie-Weiss temperature is $T^{\rm{CW}}_b=413$~K, as dictated by Eq.~(\ref{APHTZ-Tc}). This type of behavior is consistent with the one obtained within the  Kittel model \cite{Kittel:1951} for AFE systems. 

Having experimental data  for the transition temperature as well as for the temperature dependences of the characteristics depicted in fig.~\ref{APHTZ-pol-eps} is crucial for  ascertaining the values of the fitting parameters, especially $J_{11}$. 

\begin{figure*}[hbt]
	\centerline{
		\includegraphics[height=0.29\textwidth]{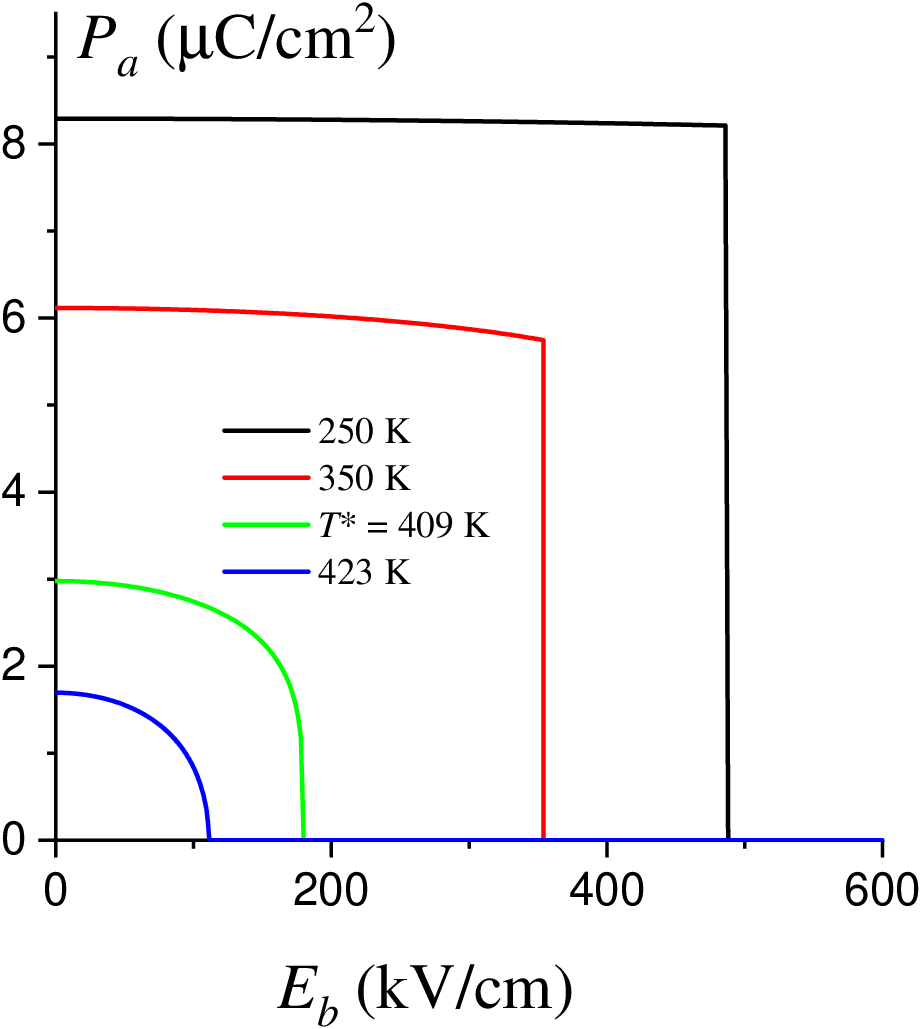}	
		\hspace{6ex}
		\includegraphics[height=0.29\textwidth]{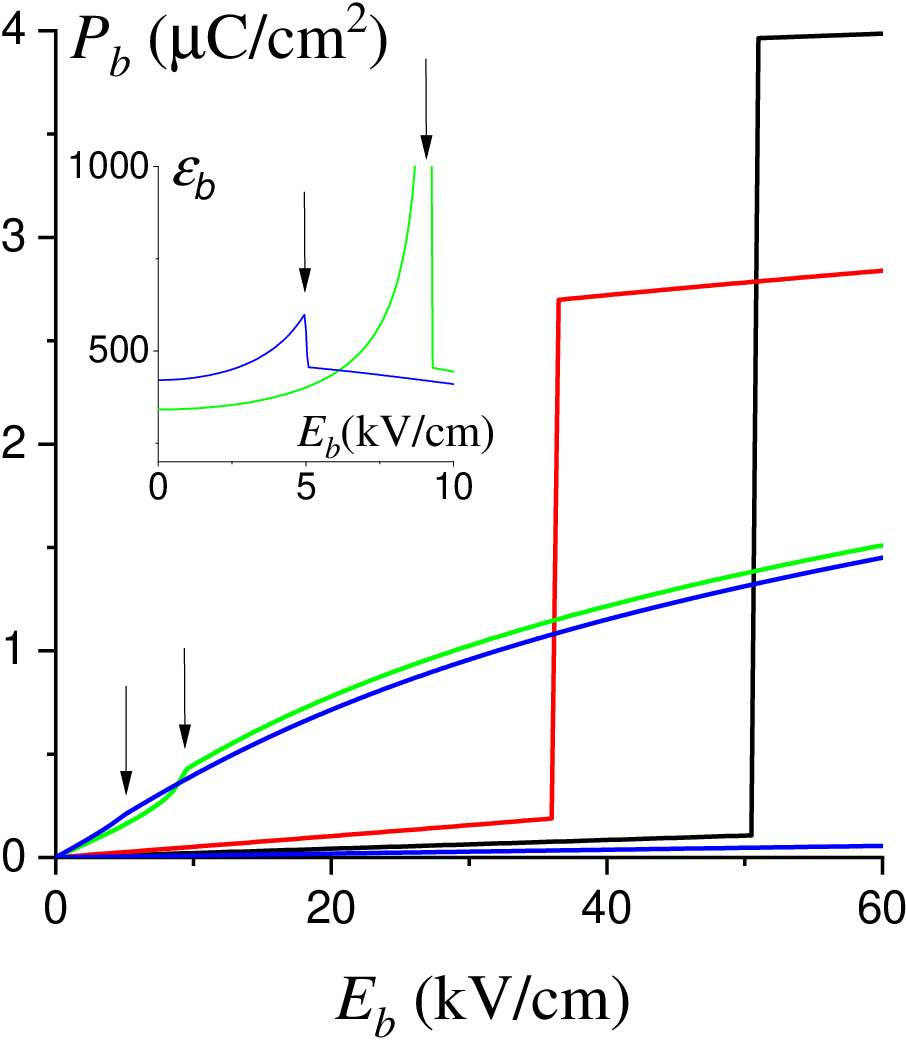}	
		\hspace{6ex}
		\includegraphics[height=0.29\textwidth]{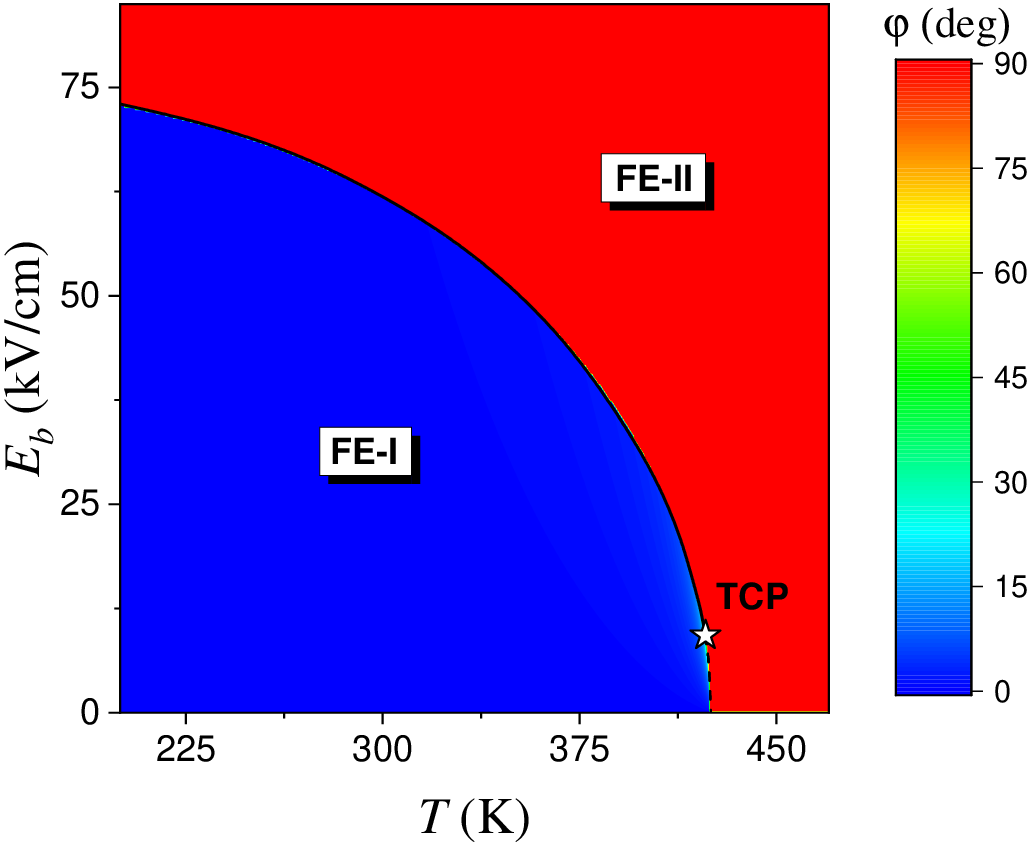}	
	}
	\caption{Dependences of the polarization components $P_a$ (left) and $P_b$ (center) of APHTZ on the field $E_b$ at different temperatures. The arrows indicates the bends in the $P_b(E_b)$ curves. The inset: the permittivity $\eps_b=\eps_0^{-1}dP_b/dE_b$ obtained by numerical differentiation of $P_b$ as a function of $E_b$. (right) The $T$-$E_b$ color contour plot of the angle $\varphi=\arctan P_b/P_a$ between the net polarization vector and the $a$ axis. Solid and dashed lines: first and second order phase transitions, respectively;  
 \faStarO:	the tricritical point $(T^*,E_b^*)$.} \label{APHTZ-switching}
\end{figure*}

As predicted, application of the field $E_b$ switches the polarization in one of the sublattices in the AFE-like manner and induces the transition from FE-I to FE-II phase. 
This process is illustrated in fig.~\ref{APHTZ-switching}.
At low temperatures the switching  is a first order phase transition, at which $P_b$ has an upward jump, whereas $P_a$
falls precisely to zero. Double AFE-type $P_b$-$E_b$ hysteresis loops are expected to be experimentally observed for the field $E_b$, as opposed to the already observed \cite{Horiuchi:2023} single FE-type $P_a$-$E_a$ ones for $E_a$.

The magnitudes of $P_a$ and $P_b$ jumps decrease with increasing temperature, until just below $T_c$, at $T^*=423$~K the transition becomes of the second order: the tricritical point occurs. At $T^*$ the left hand derivative $dP_b/dE_b$ diverges as $E_b\to E^*-0$, i.e. below the switching. Above $T^*$ only a barely discernible bend of the $P_b(E_b)$ curve can be observed; the corresponding permittivity curve $\eps_b(E_b)$, however, has a visible peak at that field (see the inset in fig.~\ref{APHTZ-switching}). Above $T_c$ there is no spontaneous polarization $P_a$ to be switched, so the field $E_b$ simply induces some non-zero $P_b$. This is  the typical behavior of an Ising type system with antipolar ordering  \cite{Bidaux:1967}.

The $T$-$E_b$ phase diagram of APHTZ overlapped by the color contour plot of the angle  $\varphi=\arctan P_b/P_a$ between the net polarization vector and the $a$ axis is presented in fig.~\ref{APHTZ-switching}. This angle is strictly 90$^\circ$ in the FE-II phase and predominantly small in the FE-I phase, except for a very narrow region near the transition line in the vicinity of the tricritical point. Thus, the switching, as expected according to the scheme in fig.~\ref{APHTZ-model}, is a 90$^\circ$ rotation of the net polarization.

 As one can see,
the temperature behavior of the switching field $E_{sw}$ follows that of the spontaneous polarization $P_a$ and order parameter  $\eta_0$. This is not accidental. Thus, for the second order transition (temperatures between $T^*$ and $T_c$) the temperature dependence of $E_{sw}$ is determined by 
the power law
\begin{equation}
\label{APHTZ-esw}
\mu_bE_{sw}\simeq J_{12}\sqrt{1-\frac{T}{T_c}}= \frac{J_{12}\eta_0}{\sqrt 3},
\end{equation}
giving an explicit connection between the switching field and the spontaneous order parameter $\eta_0$ at these temperatures. Derivation of Eq.~(\ref{APHTZ-esw}) and subsequent Eqs. (\ref{APHTZ-Eswsat}) and (\ref{APHTZ-tri}) is given in Appendix.

For the first order transitions below $T^*$ the connection between $E_{sw}$ and $\eta_0$ is just as straightforward
\begin{equation} 
\label{APHTZ-Eswsat}
\mu_b E_{sw}=\frac{J_{12}\eta_0}2.
\end{equation}
Equations (\ref{APHTZ-esw}) and (\ref{APHTZ-Eswsat}) have another obvious and intuitively clear physical meaning: the stronger is the coupling between the sublattices, the larger field is required to switch the polarization in one of them.

Finally, the temperature of the tricritical point is
\begin{equation}
	\label{APHTZ-tri}
	T^*=T_{c}\left( 1-\frac13\frac {J_{12}}{J_{11}}\right).
\end{equation}
The stronger are the intersublattice interactions $J_{12}$ as compared to the intrasublattice $J_{11}$, the farther is the tricritical point from $T_c$. For the adopted values of $J_{12}$ and $J_{11}$ it yields $T^*=423$~K, just as predicted by the direct calculations.

\begin{figure}[hbt]
	\centerline{
		\includegraphics[height=0.35\textwidth]		{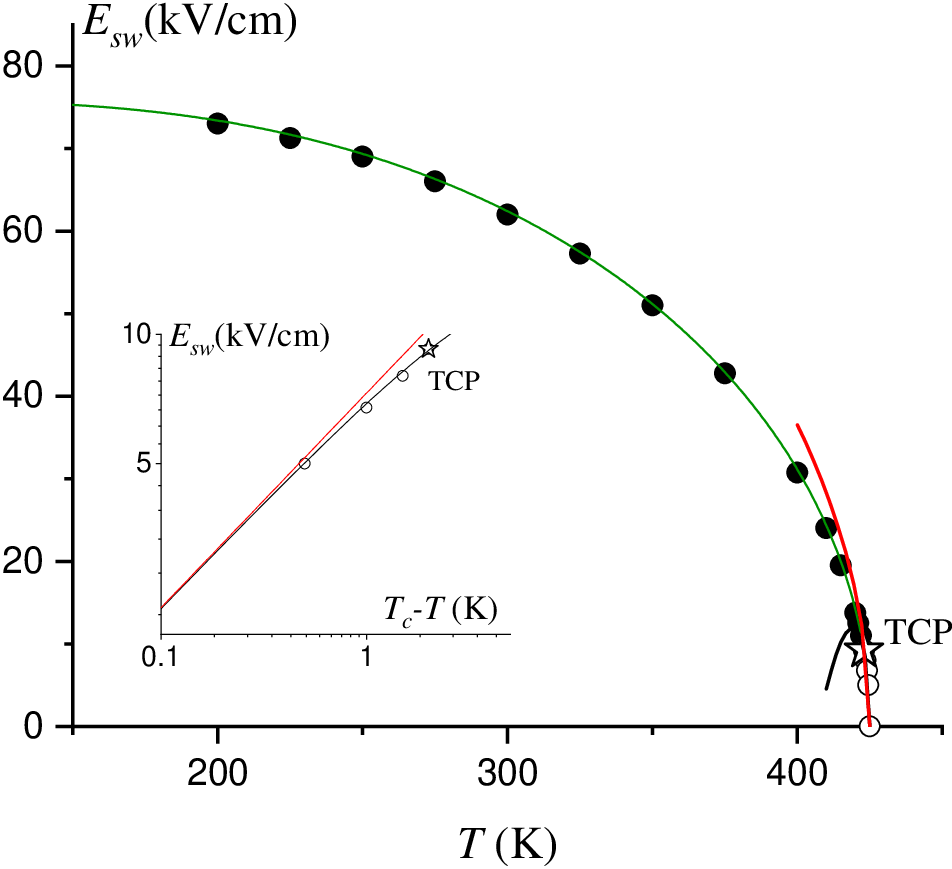}	
	}
	\caption{The temperature dependences of the switching field $E_{sw}$ in APHTZ. Symbols: direct calculations ($\Large\circ$ and $\bullet$: the second and first order phase transitions, respectively; \faStarO: the tricritical point); black lines: Eq.~(\ref{APHTZ-eswfull}); red lines: Eq.~(\ref{APHTZ-esw}); green line: Eq.~(\ref{APHTZ-Eswsat}). } \label{APHTZ-loglog}
\end{figure}

As one can see in fig.~\ref{APHTZ-loglog}, where the 
directly calculated switching fields are compared to those obtained within the presented above approximations, Eq.~(\ref{APHTZ-Eswsat}) for the first order transitions is adequate in the whole temperature range below $T^*$ up to the tricritical point. The switching field curve for the second order transitions at $T>T^*$, obtained from Eq.~(\ref{APHTZ-esw}) deviates from the exact results of Eq.~(\ref{APHTZ-eswfull}) only very slightly.


\section{MPHTZ}
\begin{figure*}[hbt]
	\centerline{\includegraphics[height=0.7\columnwidth]{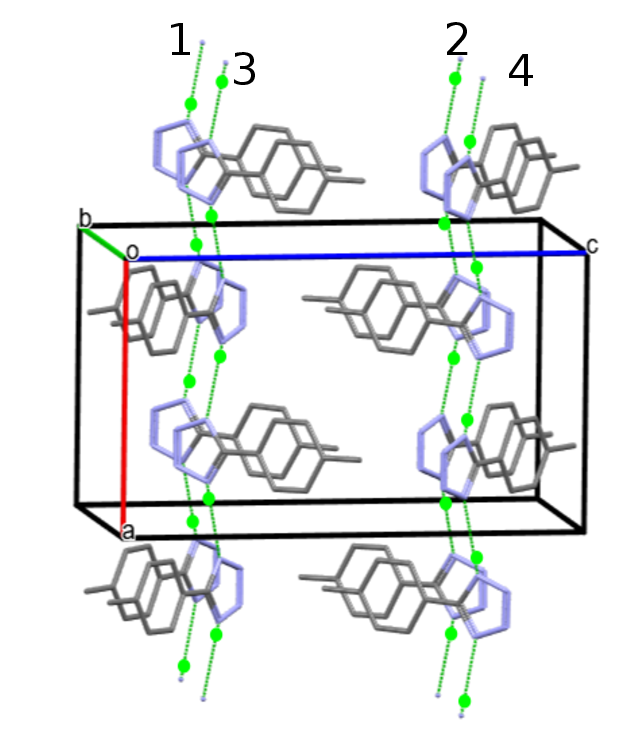}
		\hspace{10ex}
		\includegraphics[height=0.6\columnwidth]{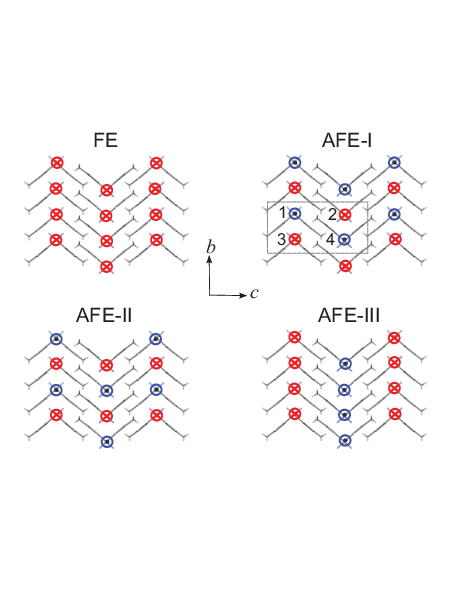}
	}
	\caption{(left) Crystal structure of MPHTZ in the AFE-I configuration, visualized \cite{Mercury} using the X-ray data of Ref.~\cite{Horiuchi:2023}. Hydrogen bonds, linking the tetrazole groups, are shown in green; $s=1,~2,~3,~4$  are the indexes of the sublattices. The off-center displacements of protons along the bonds are exaggerated.  (right) Possible FE and AFE long-range configurations of MPHTZ (drawn after Ref.~\cite{Horiuchi:2023}). Polarizations in each chain are parallel to the $a$ axis, perpendicular to the plane of the figure. Black dashed rectangle outlines the unit cell of the model.} \label{MPHTZ-struct}
\end{figure*}

The orthorhombic (space group $Pbca$) crystal structure of MPHTZ crystals is shown in fig.~\ref{MPHTZ-struct}. Unlike in APHTZ, here all hydrogen bonded chains are parallel and go along the $a$-axis without canting. It appears that there is a FE ordering within each chain and an AFE ordering of polarizations of different
chains. Directions of the chain polarizations can alternate differently: in a checkerboard manner (configuration AFE-I in fig.~\ref{MPHTZ-struct}), from layer to layer along the $b$ axis (layers parallel to the axis $c$, configuration AFE-II) or along the $c$ axis (layers parallel to $b$, configuration AFE-III). According to the DFT calculations \cite{Horiuchi:2023},  configuration AFE-I has the lowest energy.

The proposed model is naturally chosen to consist of going along the $a$ axis chains of Ising-like dipoles. The system is divided into four sublattices, as depicted in fig.~\ref{MPHTZ-struct}. The same pseudospin formalism is used, as in the previous Section, attributing the dipoles to the protons moving in the double-well potentials, to which the switchable $\pi$ bonds in the tetrazole rings and ionic displacements are rigidly tied.

The model Hamiltonian includes pairwise long-range interactions between the dipoles, as well as the energy of their interactions with the longitudinal electric field $E_a$, which is expected to flip the sublattice polarization and induce the transition to the configuration FE
\begin{equation}
\label{MPHTZ-H}
H=-\frac12\sum_{s,s'=1}^4\sum_{ff'} J_{sf \atop s'f'}\frac{\sigma_{sf}}2\frac{\sigma_{s'f'}}2 
-\mu_aE_a \sum_{s=1}^4\sum_{f} \frac{\sigma_{sf}}2 .
\end{equation}
The summations go over the four sublattices ($s=1,2,3,4$) and sites $f$ belonging to the sublattice $s$. The combined index $f$
includes the cell index, as well the index of the site, belonging to the $s$-th chain in this cell. Note that any notations introduced in this Section are relevant to MPHTZ crystals only.

For the sake of generality, we consider first the four-sublattice model with four nonequivalent  chains in each cell, which  at the proper choice of the interaction parameters can be reduced to  any of the two-sublattice configurations AFE-I, AFE-II, AFE-III from fig.~\ref{MPHTZ-struct}.  In this general model there are four distinct order parameters $\eta_s=\langle \sigma_{sf} \rangle$.
Applying the mean field approximation leads to the following Hamiltonian
\begin{equation}
H= NH'_0-\sum_{s=1}^4\sum_{f \in s } h_s \frac{\sigma_{sf}}2,
\label{MPHTZ-MFA}
\end{equation}
where $N$ is the total number of dipoles in each sublattice, 
\[
H'_0=\frac18 \sum_{ss'=1}^4 J_{ss'}\eta_{s}\eta_{s'},
\]
and
\begin{equation}
h_{s}=\frac12\sum_{s'=1}^4 J_{ss'}\eta_{s'}+\mu_aE_a.
\end{equation}
The interaction constants $\displaystyle{J_{ss'}=\sum_{ f'\in s' }J_{s f \atop s' f'}}$
have the following orthorhombic symmetry
\begin{eqnarray*}
	&&J_{ss'}=J_{s's}, \quad J_{11}=J_{22}=J_{33}=J_{44};\\
	&&J_{12}=J_{34}, \quad J_{13}=J_{24}, \quad J_{14}=J_{23}.
\end{eqnarray*}

From here the
thermodynamic potential is obtained straightforwardly
\begin{equation}
\label{MPHTZ-g}
g=\frac{G}N=H'_0
-\frac{1}{\beta}\sum_{s=1}^4\ln 2\cosh \frac{\beta h_s}{2}.
\end{equation}
The form of equations for the order parameters  $\eta_s$ is identical to that in APHTZ, Eq.~(\ref{APHTZ-ordpareq}).
Finally, the longitudinal polarization is
\begin{equation}
	\label{APHTZ-pol}
	P_{a}=\frac{\mu_a}{2v}(\eta_1+\eta_2+\eta_3+\eta_4), 
\end{equation}
$v=784~\AA^3$ is the  crystal volume per \textit{four} formula units \cite{Horiuchi:2023}.

As in the APHTZ case, it is obligatory that $J_{11}>0$ to ensure the ferroelectric ordering and absence of modulation within each separate chain (sublattice).
The interaction  constant $J_{14}$ does not bring any new essential physics, but increases the number of the model parameters. We put $J_{14}=0$. The signs of $J_{12}$ and $J_{13}$ then solely determine the character of the AFE ordering amongst the chains: $J_{12}<0$, $J_{13}<0$ yield configuration AFE-I; $J_{12}>0$, $J_{13}<0$ configuration AFE-II; and $J_{12}<0$, $J_{13}>0$ configuration AFE-III. 

Since  configuration AFE-I has been found  \cite{Horiuchi:2023} to have the lowest energy, we shall put $J_{12}<0$, $J_{13}<0$.
Under such choice, the sublattice polarizations alternate in a checkerboard pattern from one chain to the neighboring one, and the number of the order parameters reduces to two with the following symmetry: $\eta_1=\eta_4$ and $\eta_2=\eta_3$ at $E_a\neq 0$, and to one, $\eta_1=-\eta_2\equiv\eta_0$ in absence of the field.
Since in this case only the sum $J_{12}+J_{13}$ enters all the expressions, for the sake of simplicity we also take $J_{12}=J_{13}$, i.e. impose a tetragonal symmetry on the model. As usual, $\eta_A=(\eta_1-\eta_2)/2$ and $\eta_F=(\eta_1+\eta_2)/2$ can be considered as auxiliary AFE and FE order parameters, respectively.

The thermodynamic potential (per \textit{two} molecules) can now be rewritten in exactly the same form as Eq.~(\ref{APHTZ-g}), with
\[
H_0=\frac18[J_{11}\eta_1^2+4J_{12}\eta_1\eta_2 +J_{11}\eta_2^2],
\]
and
\begin{eqnarray}
	&& h_{1}=\frac{J_{11}\eta_1+2J_{12}\eta_2}{2}+\mu_aE_a, \nonumber\\
	&& h_{2}=\frac{2J_{12}\eta_1+J_{11}\eta_2}{2}+\mu_aE_a.
	\label{MPHTZ-h}
\end{eqnarray}
The proposed model for MPHTZ is thus reduced to the traditional two-sublattice AFE Ising model, whose thermodynamics has been thoroughly studied yet in Ref.~\cite{Bidaux:1967} and is well known.

By differentiating the polarization $P_a=2\mu_a\eta_F/v$, we find the longitudinal permittivity at zero bias field
 \begin{equation}
 \eps_a=\frac{\beta\mu_a^2}{\eps_0v\bar\varphi_a},
 	\end{equation}
 	where 
 	\[
 	\bar\varphi_{a}=\frac{1}{1-\eta_0^2}-\frac{\beta(J_{11}+2J_{12})}{4}.
 	\]
 It is expected to remain finite at all temperatures.
 The Curie-Weiss law is obeyed in the paraelectric phase, with the Curie-Weiss temperature 
 	\begin{equation}
\label{MPHTZ-CW}
T^{\rm{CW}}=\frac{J_{11}+2J_{12}}{4k_{\rm B}}.
 	\end{equation}

The expressions for the predicted second order phase transition temperature at zero bias field 
\begin{equation}
\label{MPHTZ-TN}
T_{\rm N}=\frac{J_{11}-2J_{12}}{4k_{\rm B}},
\end{equation}
the temperature dependence of the field of switching to the FE phase 
\begin{equation} 
\label{MPHTZ-Eswsat}
\mu_a E_{sw}=-{J_{12}\eta_0}
\end{equation}
for the first order transitions at temperatures below $T^*$ and
\begin{equation}
\label{MPHTZ-esw}
\mu_a E_{sw}\simeq -2J_{12}\sqrt{1-\frac{T}{T_{\rm N}}},
\end{equation}
for the second order transitions  above $T^*$, where
the temperature of the tricritical point is
\begin{equation}
\label{MPHTZ-tri}
T^*=T_{\rm N}\left( 1+\frac23\frac {J_{12}}{J_{11}}\right),
\end{equation}
have been derived in Appendix. Equations~(\ref{MPHTZ-esw}) and (\ref{MPHTZ-tri})  were obtained yet in Ref.~\cite{Bidaux:1967}.

Additionally, the same power law
\begin{equation}
\label{MPHTZsqrt}
\eta_0=\sqrt{3\frac{T_{\rm N}-T} {T_{\rm N}}
}
\end{equation}
and exponential 
\begin{equation}
\label{MPHTZetaapproxsat}
\eta_0=1-2\exp\left(\frac{-2T_{\rm N}}{T}\right)\end{equation}
approximations of the spontaneous order parameter at temperatures close to the phase transition and to saturation are obtained
for the case of the antiferroelectric MPHTZ, as in the previous Section for APHTZ.

\begin{figure*}[hbt]
	\centerline{
		\includegraphics[height=0.27\textwidth]{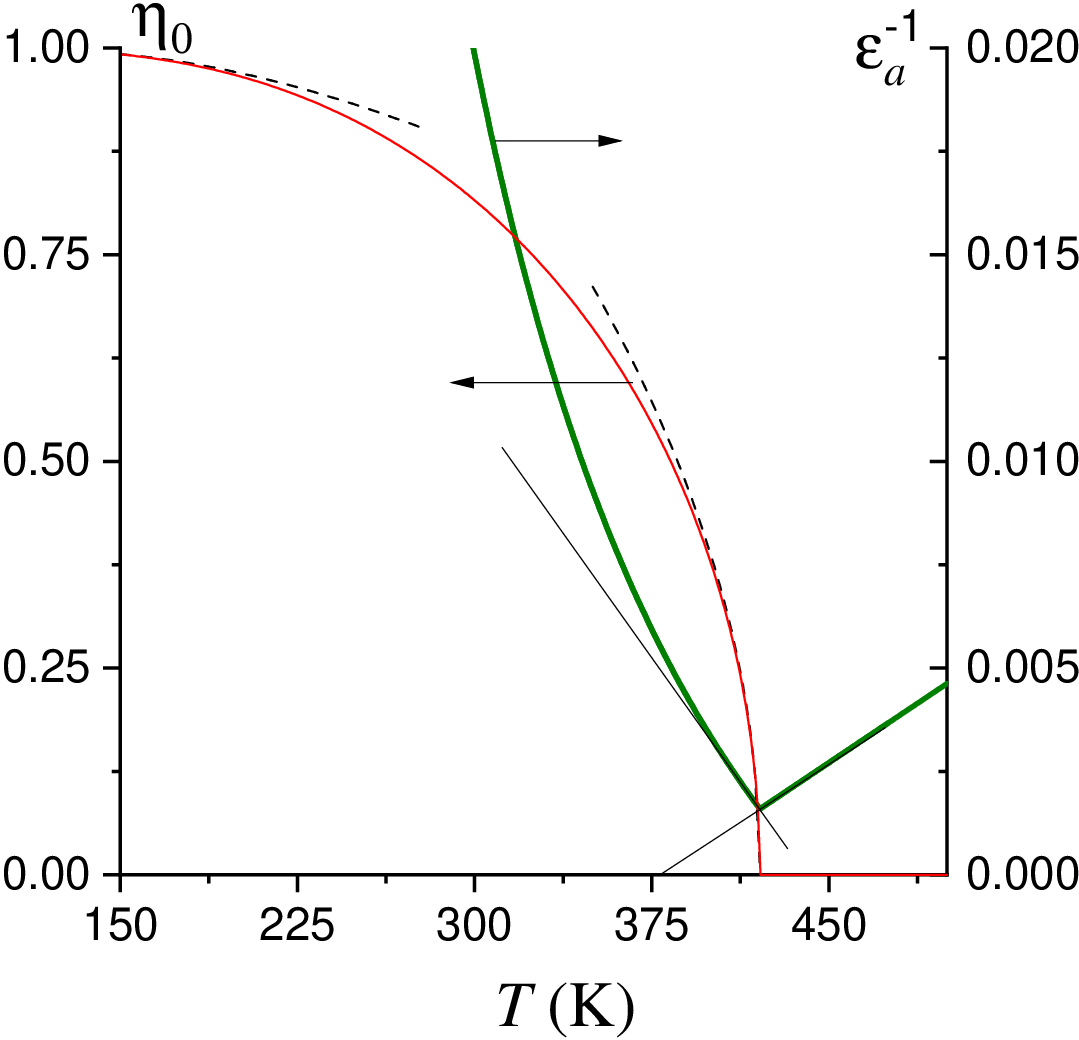}
		\hspace{8ex}
		\includegraphics[height=0.27\textwidth]{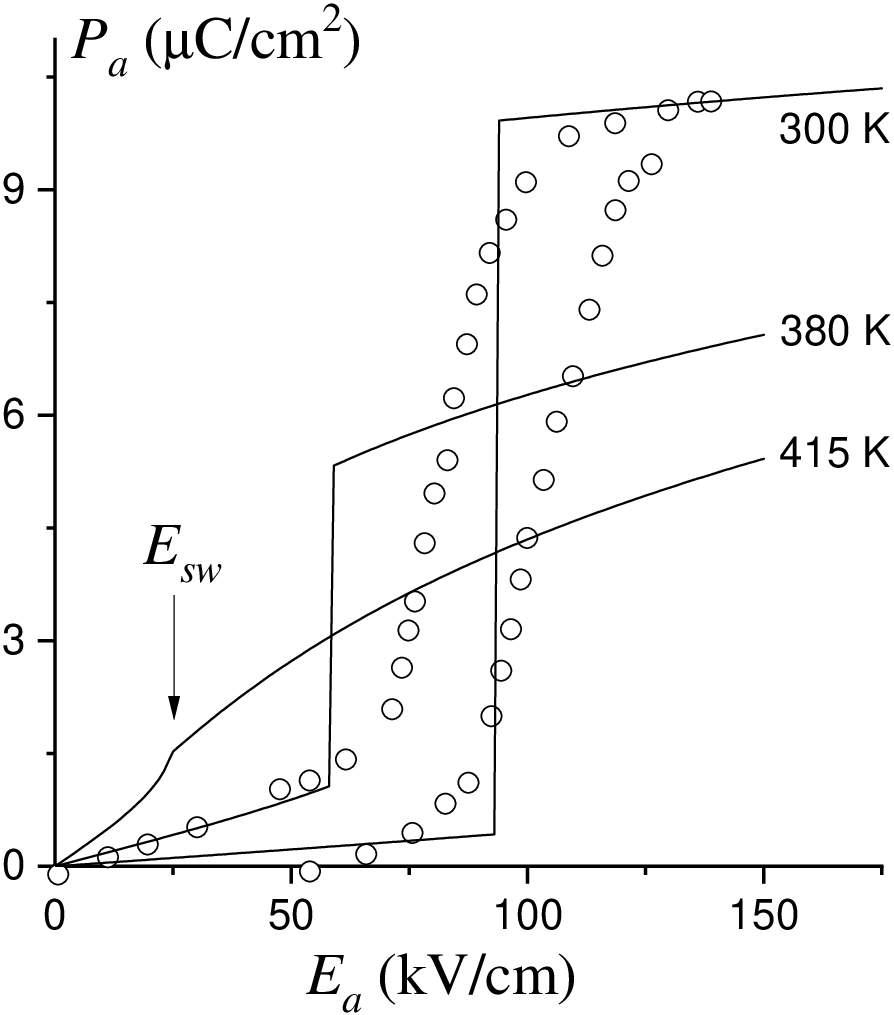}	
		\hspace{8ex}
		\includegraphics[height=0.27\textwidth]{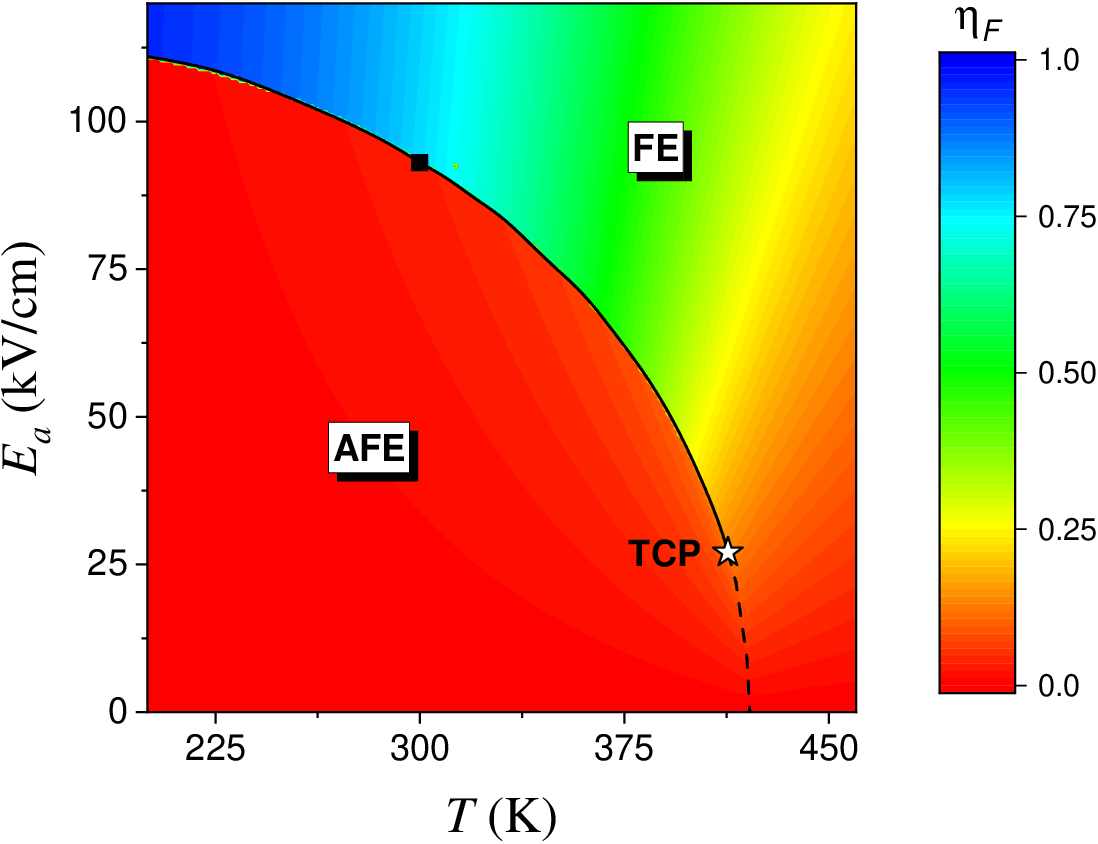}	
	}
	\caption{(left) Dependences of the spontaneous order parameter $\eta_0$, inverse dielectric permittivity $\eps^{-1}_a$ of MPHTZ on temperature.  Solid lines: the theory; dashed lines: $\eta_0$ approximated according to Eqs. (\ref{MPHTZsqrt}) and (\ref{MPHTZetaapproxsat}).  (center) Dependences of the net  polarization $P_a$ on the field $E_a$ at different  temperatures. $\circ$:  experimental data \cite{Horiuchi:2023} for 300~K. (right) The $T$-$E_a$ color contour plot of the FE order parameter $\eta_F$. Solid and dashed lines: the first and second order phase transitions, respectively; \faStarO: the tricritical point $(T^*,E_a^*)$; $\blacksquare$:  experimental point \cite{Horiuchi:2023}.} \label{MPHTZ-switching}
\end{figure*}

The fitting procedure for MPHTZ is similar to that for APHTZ.
The parameters $\mu_a=5\cdot10^{-29}$~C m and $J_{12}/k_{\rm B}=-41.6$~K are chosen to obtain the best fit to the experimental dependence \cite{Horiuchi:2023} of polarization $P_a$ on the field $E_a$ at 300~K, in particular, to the values of the switching field and polarization above the switching. The parameter $J_{11}/k_{\rm B}=1600$~K is then chosen to set $T_{\rm N}$  above the experimentally proven \cite{Horiuchi:2023} minimal stability limit of the AFE phase, 410~K. At the adopted values of $J_{11}$ and $J_{12}$, the theoretical $T_{\rm N}$ is 421~K. Again, $|J_{12}|\ll J_{11}$, i.e. the coupling between the sublattices is much weaker than within them.

Alternatively, the intersublattice interaction parameter $J_{12}$ can be determined according to the same scheme as for APHTZ, by equating the energy difference between the fully ordered FE  ($\eta_1 \to 1$, $\eta_2\to 1$)  and AFE-I ($\eta_1 \to 1$, $\eta_2\to -1$) states  found from the model thermodynamic potential to $\Delta E= 1.78$~meV/molecule \cite{Horiuchi:2023}, obtained by the DFT calculations
\[
\frac{g^{\rm{FE}}-g^{\rm{AFE{-}I}}}2=-\frac{J_{12}}2 =\Delta E.\]
This yields basically the same value of $J_{12}/k_{\rm B}=-41.3$~K, as that found by fitting to the experimental $P_a(E_a)$ dependence. This agreement between the two independent estimates of $J_{12}$ is an important evidence in favor of the model validity.

The temperature dependence of the spontaneous order parameter $\eta_0$ is presented in fig.~\ref{MPHTZ-switching}. 
The power law and exponential approximations (\ref{MPHTZsqrt}) and (\ref{MPHTZetaapproxsat}) are adequate down to $T_{\rm N}-T\simeq 25$~K and up to 210~K, respectively. The temperature curve of the inverse permittivity $\eps_a^{-1}(T)$  is as  expected from the Kittel model: it
follows the Curie-Weiss law and remains finite at the transition point. The  paraelectric Curie-Weiss temperature is $T^{\rm{CW}}=379$~K, as follows from Eq.~(\ref{MPHTZ-CW}). 

The field-induced switching from AFE to FE phase is typical for uniaxial antiferroelectrics \cite{Bidaux:1967,toledano:16}.
At lower temperatures the switching is a first order phase transition. With increasing temperature the polarization jump decreases and eventually transforms to a bend in the $P_a(E_a)$ curve (c.f. the $P_b(E_b)$ curves, fig.~\ref{APHTZ-switching}), the bevahior indicative of a presence of the tricritical point.

This agrees with the $T$-$E_a$ phase diagram presented in fig.~\ref{MPHTZ-switching}. This diagram is topologically identical to that of APHTZ for the transverse field $E_b$ (fig.~\ref{APHTZ-switching}). However, here the tricritical point occurs further from $T_{\rm N}$, at 413~K, which agrees with Eq.~(\ref{MPHTZ-tri}),  and the values of the switching fields are also overall larger than those obtained for APHTZ. These differences are attributed to the weaker intersublattice interactions in APHTZ, inferred from the smaller energy difference between the states below and above the switching \cite{Horiuchi:2023}, and follow directly from Eqs.~(\ref{APHTZ-esw})-(\ref{APHTZ-tri}) and (\ref{MPHTZ-Eswsat})-(\ref{MPHTZ-tri}). 

\begin{figure}[hbt]
	\centerline{
		\includegraphics[height=0.35\textwidth]		{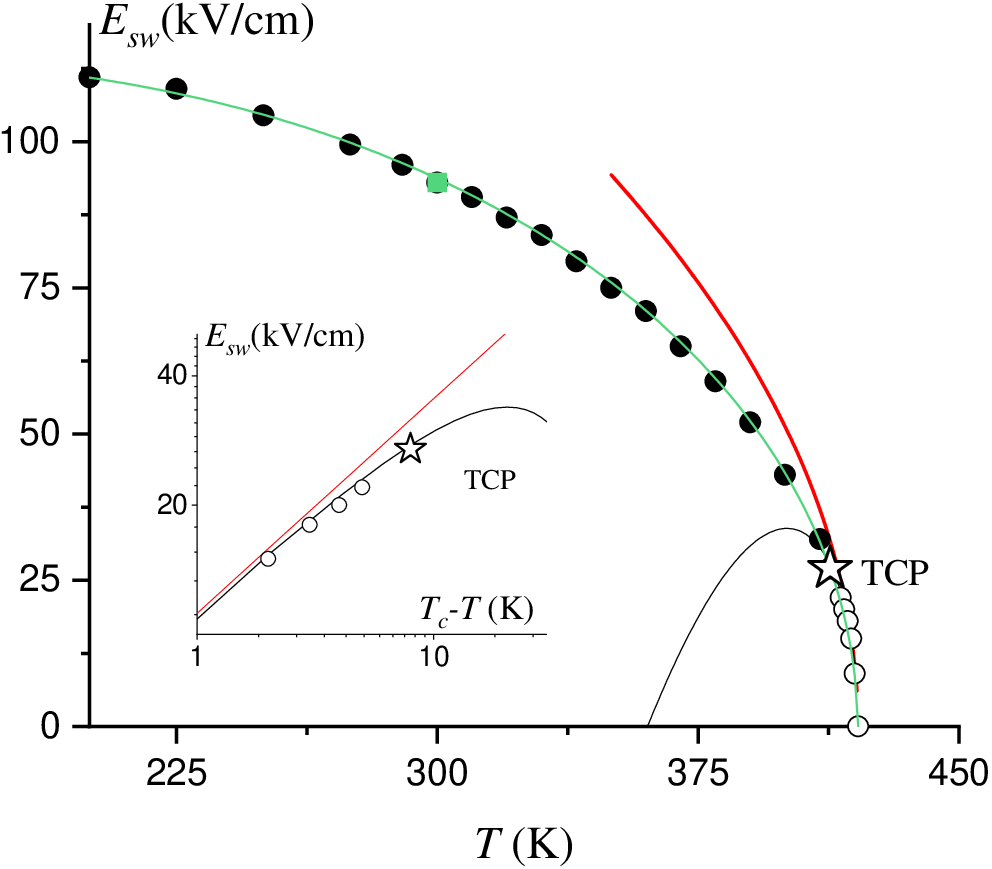}	
	}
	\caption{The temperature dependences of the switching field $E_{sw}$ in MPHTZ. Symbols: direct calculations ($\Large\circ$ and $\bullet$: the second and first order phase transitions, respectively; \faStarO: the tricritical point); black lines: Eq.~(\ref{MPHTZ-eswfull}); red lines: Eq.~(\ref{MPHTZ-esw}); green line: Eq.~(\ref{MPHTZ-Eswsat}). {\color{Green}  $\blacksquare$}: the experimental point \cite{Horiuchi:2023}.} \label{MPHTZ-loglog}
\end{figure}

Adequacy of the approximations (\ref{MPHTZ-Eswsat})-(\ref{MPHTZ-tri}) is verified in fig.~\ref{MPHTZ-loglog}.
All conclusions drawn in the previous Section from the similar graph of the switching field for APHTZ, Fig.~\ref{APHTZ-loglog}, remain valid for MPHTZ as well. The switching field is indeed proportional to the spontaneous order parameter $\eta_0$ and to the intersublattice interaction parameter $J_{12}$ at all temperatures, in accordance with Eqs.~(\ref{MPHTZ-esw}) and (\ref{MPHTZ-tri}).

Even though the present theory considers only the equilibrium states of the system, and therefore, there is no hysteresis at the first order phase transition switching, we still can draw some conclusions about the  stored and recoverable energy densities for the studied crystals, the primary characteristics of a system that potentially can be used for energy storage. By definition, these densities are given by the areas between the polarization axis and, respectively, right/left branches of the $P_a(E_a)$ hysteresis loop at $E_a>0$ (forward/backward transitions). 
The area $W=\int E_adP_a$ between the theoretical $P_a(E_a)$
curve and the polarization axis  can then serve as the lower and upper limit estimates of the actual stored and recoverable energy densities.

It is obvious from Fig.~\ref{MPHTZ-switching} that at the same value of the maximal applied field ($E_{max}>E_{sw}$), this area is the largest if there is no significant increase of polarization prior to and after the switching, i.e. the corresponding dielectric permittivity is small, and the switching is just polarization flipping in one of the sublattices. In other words, the working temperature should be far from $T_c$ and far from $T^*$, where the permittivity diverges (see the inset to fig.~\ref{APHTZ-switching}). We then estimate the optimal recoverable energy density by approximating the $P_a(E_a)$ curve by a step-like increase of polarization from zero to $\Delta P_a=2\mu_a\eta_0/v$  at $E_{sw}$. Using  Eq.~(\ref{MPHTZ-Eswsat}) we obtain that
\begin{equation}
W = E_{sw} \Delta P_a=-\frac{2J_{12}\eta_0^2}v.
\end{equation}
With $\eta_0=0.815$ at 300~K it yields $W=0.966$~J/cm$^3$ for MPHTZ, in a good agreement with the experimental \cite{Horiuchi:2023} stored energy density value 1.044~J/cm$^3$.

Counterintuitively, $W$ does not depend on the value of the dipole moment $\mu_a$, which is canceled out in the $E_{sw}\Delta P_a $ product. Larger intersublattice interactions $J_{12}$ increase $W$, provided that the ratio $J_{12}/J_{11}$ is still small enough for the tricritical point to be far from the working temperature [see Eq. (\ref{MPHTZ-tri})].
The more the  system is ordered  at zero field (the closer $|\eta_0|$ is to 1), the larger is $W$. Maximal possible $W$ is therefore $W_{max}=-2J_{12}/v$ or 1.45~J/cm$^3$ for a fully ordered MPHTZ.

\section{Concluding remarks and outlook}

We propose pseudospin models, which reproduce the essential structure of the ordering subsystem in the two crystals of the phenyltetrazole family, APHTZ and MPHTZ, and, hopefully, can qualitatively and quantitatively correctly describe their temperature and electric field behavior. A good agreement with the available experimental data is obtained.

However, these data are still not sufficient to make definite conclusions about the adequacy of the developed approach. First, the obtained critical behavior of the models (the second order phase transition, the $\sqrt{T_c-T}$ dependence of the order parameter, the Curie-Weiss law for the permittivities, the $T$-$E$ phase diagrams with a tricritical point) is as expected for the Ising-type systems treated in the mean field approximation. That needs to be confirmed experimentally.
Second, though the number of the used free parameters is absolutely minimal, there is still a certain degree of uncertainty (in particular for the values of the intrasublattice interaction parameter $J_{11}$), caused by the severe lack of experimental data to base the fitting on. 
Therefore, measurements of the temperature dependences of spontaneous polarization, dielectric permittivities, polarization switching, eventual determination of temperatures and order of the phase transitions to the paraelectric phase are absolutely necessary in order to test the model validity and ascertain the values of the free parameters. It may also transpire that further sophistications of the models are required, like employing the two-particle cluster approximation for the intrachain correlations.

The processes of polarization switching by external electric field are studied in detail.  MPHTZ behaves like a classical uniaxial AFE: the field-induced net polarization is confined to this axis, and the classical switching with double $P$-$E$ hysteresis loops is observed \cite{Horiuchi:2023}. APHTZ, a canted FE, is different: it behaves like a ferroelectric, when spontaneous or associated with the longitudinal field $E_a$ applied along the $a$ axis characteristics are concerned, and like an antiferroelectric, when associated with transverse field $E_b$, perpendicular to the axis of spontaneous polarization, characteristics are concerned. Flipping one of sublattice polarizations by 180$^\circ$ by the transverse field leads to a 90$^\circ$ rotation of the net polarization.  In addition to the single FE loop observed \cite{Horiuchi:2023} for the longitudinal field $E_a$, one should also observe a double loop for the transverse field $E_b$. To investigate the effect of changing the external field direction, i.e. combining the longitudinal $E_a$ and transverse $E_b$ bias fields, on the polarization rotation in APHTZ could be particularly interesting.

MPHTZ crystals may also present an academically interesting case as far as the electrocaloric effect  in them is concerned. Commonly, in FEs it is positive, $dT/dE>0$, whereas in AFEs it can be negative, $dT/dE<0$, or positive in different field or temperature domains \cite{grunebohm:18,Moina:2023}. It would be interesting to determine the sign of the electrocaloric effect in MPHTZ, where the FE ordering along the $a$ axis coexists with the AFE ordering along the $b$ axis, and the  situation can be reversed by the field $E_b$.

The parent PHTZ compound, identified  as a mixture of two different AFE phases with close energies \cite{Horiuchi:2023}, is intriguing. Judging by the shape of the double hysteresis loops in it, the switching fields of the two phases must also be very close. The present model approach can explain that, as it appears that these two phases will be indistinguishable in the mean field approximation for the interchain interactions, thus having identical energies and switching fields.
More detailed calculations are in order.

\section*{Appendix}

\begin{figure}[hbt]
	\centerline{
		\includegraphics[height=0.33\textwidth]{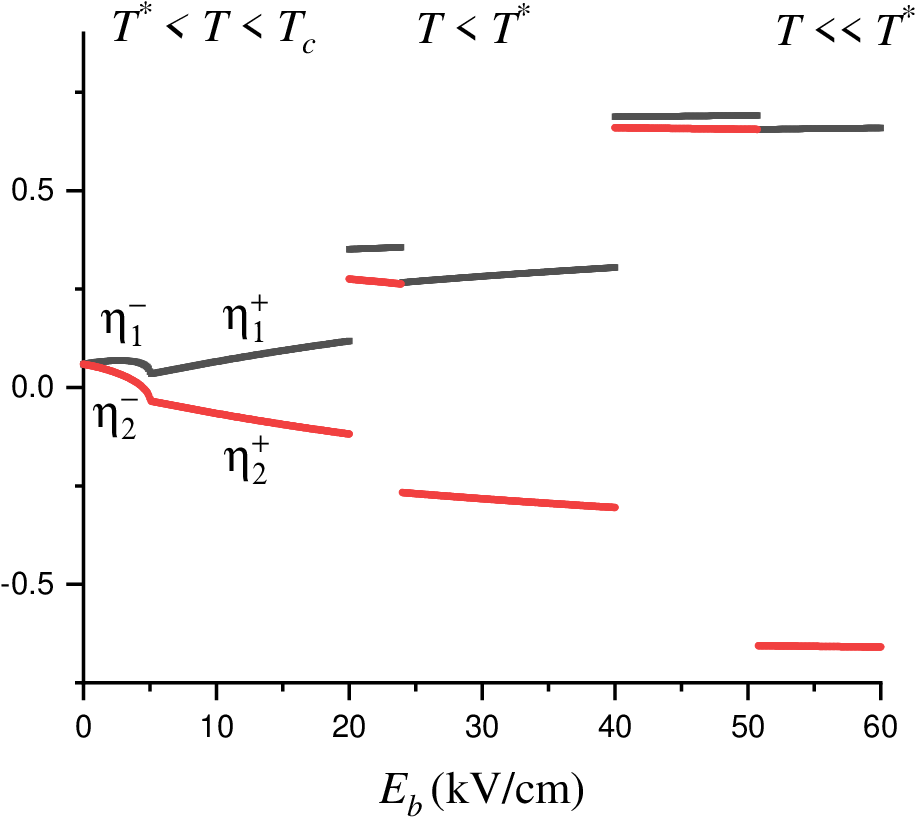}		}
	\caption{The field dependences of the order parameters in APHTZ at different temperatures: 424.5~K ($T^*<T <T_c$), 410~K ($T <T^*$), and 350~K ($T \ll T^*$). } \label{APHTZ-eta}
\end{figure}

The typical dependences of the order parameters in APHTZ in the vicinity of the switching induced by the electric field $E_b$ are depicted in fig.~\ref{APHTZ-eta}. Note that just above the transition $\eta_1^+=-\eta_2^+$. Using Eqs.~(\ref{APHTZ-ordpareq}) it can also be shown that in the linear over $E_b$ approximation, which  as seen in fig.~\ref{APHTZ-eta} is well appropriate for $T<T^*$, we have
\begin{equation}
\label{etapluseta}
\eta_1^-+\eta_2^-=2\eta_0,
\end{equation}
where $\eta_0$ is the spontaneous order parameter in the FE-I phase.

In order to obtain the temperature dependence of the switching field for the first order transitions (below $T^*$), one has to equate the thermodynamic potentials, Eq.~(\ref{APHTZ-g}), for the two phases just below and above the transition, i.e. at $\eta_1=\eta_1^-$, $\eta_2=\eta_2^-$ and at $\eta_1=-\eta_2=\eta_1^+$:
\[
-\frac 1\beta \ln \frac {\cosh (\beta h_1^- /2) \cosh (\beta h_2^- /2) }{\cosh^2 (\beta h_1^+ /2) }=H_0^+-H_0^-,
\]
where $H_0$ and $h_{1,2}$ are given by Eqs.~(\ref{APHTZH0}) and 
(\ref{APHTZh1h2}), respectively.
In a linear over $E_b$ approximation we obtain that
\[
\mu_bE_{sw}=2\frac{H_0^+-H_0^-+\beta^{-1}\ln\frac{\cosh a\cosh b}{\cosh^2 c}}{2\tanh c-\tanh a +\tanh b},
\]
where
$a=\beta (J_{11}\eta_1^-+J_{12}\eta_2^-)/4$, $b=\beta (J_{12}\eta_1^-+J_{11}\eta_2^-)/4$, and $c=\beta (J_{11}-J_{12})\eta_1^+/4$.
At small deviations of the order parameters from 1 and using Eqs.~(\ref{etaapproxsat}) and (\ref{etapluseta}), it leads to the following equation
\[
\mu_bE_{sw}=\frac{J_{12}+2\beta^{-1}(1-\eta_0)[1-\exp(\beta J_{12})]}{2[1-(1-\eta_0)[1-\exp(\beta J_{12})]]}.
\]
Expanding it in series and retaining up to the linear terms in $\beta J_{12}$ and $1-\eta_0$, we obtain  a remarkably simple final expression for the switching field, Eq.~(\ref{APHTZ-Eswsat}). As one can see in fig.~\ref{APHTZ-loglog}, despite all the used approximations, $E_{sw}$ given by this expression agrees with the results of direct calculations surprisingly well.

Curiously, Eq.~(\ref{APHTZ-Eswsat}) can be obtained in a much simpler way, if one uses an empirical fact that at the switching point $\eta_2^-\simeq \eta_1^+$ at low temperatures (see fig.~\ref{APHTZ-eta}). Equating $h_{2}^-$ to $h_{1}^+$ and taking into account Eq.~(\ref{etapluseta}), one immediately arrives at Eq.~(\ref{APHTZ-Eswsat}).

To obtain the temperature dependence of the switching field above $T^*$, we use the condition that the derivative $dP_a/dE_a$ diverges at the second order phase transition point. 
Taking into account the fact that at the transition $\eta_1=-\eta_2\equiv\eta_{sw}$ (see fig.~\ref{APHTZ-eta}) and
calculating the derivative at $E_b\neq 0$, we obtain that 
\[
\eta_{sw}=\sqrt{1-\frac T{T_c}}.
\]
Substituting it into Eq.~(\ref{APHTZ-ordpareq}), we find an exact expression for the switching field above $T^*$
\begin{equation}
\label{APHTZ-eswfull}
\frac{\mu _b E_{sw}}{k_{\rm B}}=2T\arctanh\sqrt{1-\frac{T}{T_c}}-\left(2T_c-\frac{J_{12}}{k_{\rm B}}\right)\sqrt{1-\frac{T}{T_c}}.
\end{equation}
At temperatures close to $T_c$ it can be approximated by Eq.~(\ref{APHTZ-esw}).

The temperature of the tricritical point  $T^*$ can be found  by following the scheme suggested in \cite{Bidaux:1967}. We introduce auxiliary order parameters $\eta_{\rm I}=(\eta_1+\eta_2)/2$ and $\eta_{\rm{II}}=(\eta_1-\eta_2)/2$. 
and rewrite 
Eqs.~(\ref{APHTZ-ordpareq}) as
\begin{eqnarray}
	&& \frac{2\eta_{\rm{I}}}{1+\eta_{\rm{I}}^2-\eta_{\rm{II}}^2}=\tanh\left[ \beta\mu_b E_b+\left(2\frac{T_c}{T}-\beta J_{12}\right)\eta_{\rm{I}}\right], \nonumber\\
	\label{Bidaux}
	&&\frac{2\eta_{\rm{II}}}{1-\eta_{\rm{I}}^2+\eta_{\rm{II}}^2}=\tanh \left[2\frac{T_c}{T}\eta_{\rm{II}}\right].
\end{eqnarray}
Since the tricritical point ($T^*, E_b^*$) is the terminating point of the second order transition line, at this point  $\eta_{\rm{II}}=0$ and $\eta_{\rm{I}}=\eta_{sw}$.  We then expand Eqs.~(\ref{Bidaux}) at $T=T^*$ in the vicinity of $E_b^*$  with $\Delta E = E_b-E_b^*<0$, $\eta_{\rm{II}}$, and $\Delta \eta=\eta_{\rm{I}}-\eta_{sw}$ as small parameters. The coordinates of the tricritical point are found from the condition that the left hand derivative $d\eta_{\rm{II}}/dE_b$ diverges, or that $\Delta E/\Delta \eta=0$. Some lengthy algebra leads to yet another remarkably simple expression, Eq.~(\ref{APHTZ-tri}).

\begin{figure}[hbt]
	\centerline{
		\includegraphics[height=0.33\textwidth]{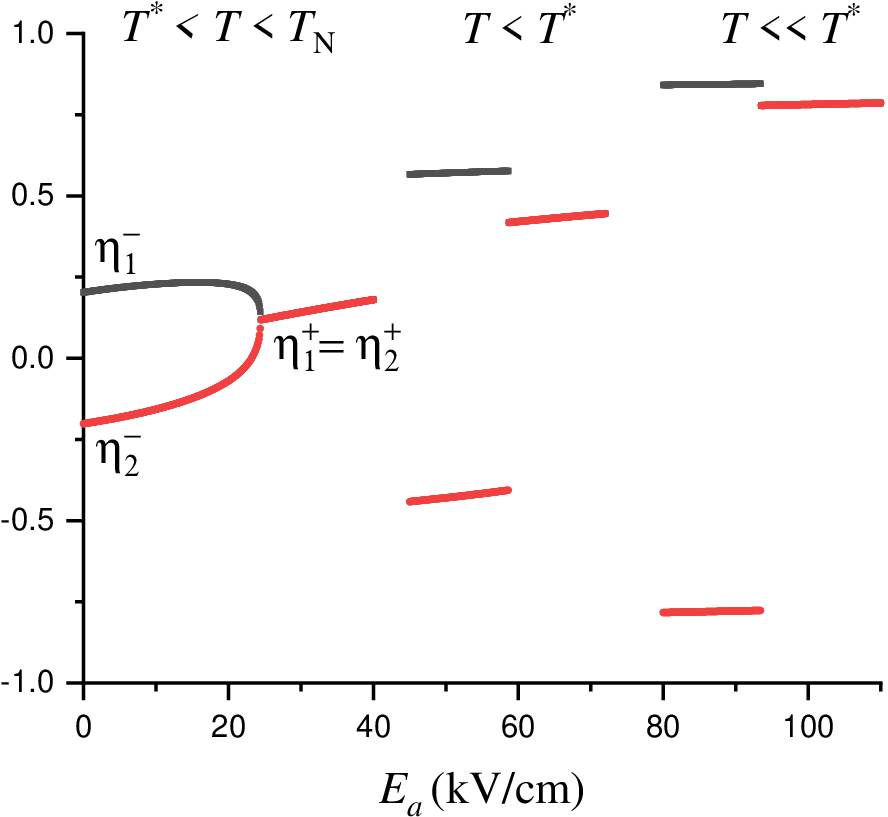}	
	}
	\caption{The field dependences of the order parameters in MPHTZ at different temperatures: 415~K ($T^*<T <T_{\rm N}$), 380~K ($T <T^*$), and 300~K ($T \ll T^*$). } \label{MPHTZ-eta}
\end{figure}

The typical dependence of the order parameters in MPHTZ in the vicinity of the switching  is depicted in fig.~\ref{MPHTZ-eta}.
As one can see, it is different from that in APHTZ. Above the transition $\eta_F=\eta_1^+=\eta_2^+$, and it can be shown that
\begin{equation}
	\label{etamineta}
	\eta_A^-=\eta_1^--\eta_2^-=2\eta_0
\end{equation}
is obeyed instead of Eq.~(\ref{etapluseta}).

To find the values of the FE order parameter  at the second-order transition switching $\eta_F^{sw}$, we assume that in addition to the bias $E_a$, a hypothetical measuring staggered field $E_{st}$ is applied, so that $h_1\to h_1+\mu_{st}E_{st}$, and  $h_2\to h_2-\mu_{st}E_{st}$ ($h_i$ are given in Eq.~(\ref{MPHTZ-h})). For the point, where the linear response of the AFE order parameter $\eta_A$ to the field $E_{st}$ diverges, one gets 
\begin{equation}
	\eta_F^{sw}=\sqrt{1-\frac{T}{T_{\rm N}}},
\end{equation}
where the transition temperature  $T_{\rm N}$ at zero  bias field, at which  $\eta_F^{sw}=0$, is given by Eq.~(\ref{MPHTZ-TN}).

In the same manner as described above for the case of APHTZ, we obtain
\begin{equation}
	\label{MPHTZ-eswfull}
	\frac{\mu_a  E_{sw}}{k_{\rm B}}=2T\arctanh\sqrt{1-\frac{T}{T_{\rm N}}}-2\left(T_{\rm N}+\frac{J_{12}}{k_{\rm B}}\right)\sqrt{1-\frac{T}{T_{\rm N}}},
\end{equation}
instead of Eq.~(\ref{APHTZ-eswfull}) for the switching fields above the tricritical point, Eq.~(\ref{MPHTZ-esw}) as its approximation, as well as expression~(\ref{MPHTZ-tri}) for the temperature of the tricritical point and Eq.~(\ref{MPHTZ-esw}) for the switching fields below it.

\end{document}